\documentclass{llncs}
\usepackage{makeidx}  
\usepackage{graphicx}  
\graphicspath{{./figures/}}
\usepackage{verbatim}
\usepackage{color}
\usepackage{url}

\begin{document}
\newcommand{\squishlist}{
	\begin{itemize}
 }
 \newcommand{\squishend}{
 \end{itemize}
 }
\title{Extracting, Transforming and Archiving Scientific Data}
\titlerunning{ETA for Scientific Data}  
%
\author{Daniel Lemire\inst{1} \and Andre Vellino\inst{2}}

\authorrunning{Daniel Lemire, Andre Vellino} 
%
\tocauthor{Daniel Lemire and Andre Vellino}
\institute{Universit\'e du Qu\'ebec \`a Montr\'eal,\\
100 Sherbrooke West\\
Montreal, QC, H2X 3P2 Canada\\
\email{lemire@gmail.com}\\
\and
CISTI - National Research Council of Canada\\
Building M-55, 1200 Montreal Road\\
Ottawa, ON K1A 0S2 Canada\\
\email{andre.vellino@nrc.ca}
}

\maketitle

\begin{abstract}
It is becoming common to archive research datasets
that are not only large
but also numerous. In addition, their corresponding
metadata and the software required to
analyse or display them need to be archived. Yet the manual curation of research data can be
difficult and expensive, particularly in very large digital repositories,
hence the importance of models and tools for automating digital curation tasks.
The automation of these tasks faces three major
challenges: (1)~research data and data sources are highly heterogeneous, 
(2)~future research needs are difficult to anticipate, (3)~data is hard to
index. To address these problems, we propose the Extract, Transform and
Archive (ETA) model for managing and mechanizing the curation of research data.
Specifically, we propose a scalable strategy for addressing the research-data
problem, ranging from the extraction of legacy data to its long-term storage.
We review some existing solutions and propose novel avenues of research.

\end{abstract}

\section{Introduction}

The conventional workflow model of science---whereby the scientist proposes a
hypothesis, devises a series of experiments, performs the experiments,
generates data and produces a publication in a peer-reviewed journal---is no
longer adequate to characterize present-day scientific endeavours. First, a
significant amount of scientific research is devoted to experimental design,
data-collecting, developing increasingly precise measurement techniques and
managing the acquired data. Furthermore, researchers today have an increasing
ability to share resources and methods and a greater need to handle large
volumes of data.  They also have more opportunity to collaborate across a
variety of disciplines and have a greater diversity of channels for
disseminating results, data and software beyond conventional publication
channels.

Hence there is a need for tools that automate the curation processes
beyond merely storing and archiving large volumes of research data. They
also need to enable data reuse, interoperability and discovery.
This challenge is especially difficult
because research communities differ so widely in their needs and
practices~\cite{1816173} that universally applicable conventions are
impossible to establish. Furthermore, to create complete data archives we must also be able
to extract data from previously published ``backfiles'' whose legacy data
content may not have ever been managed, curated or archived at all, let alone
with discovery, reuse and repurposing in mind.

To achieve these objectives, we present a new data-management model for digital
libraries which addresses the problems of large scale automation of extraction,
transformation and archiving (ETA) of scientific research data. 
Our proposal is founded on a mature model --- Extract Transform Load (ETL) --- that has been 
developed for business data
warehousing~\cite{Inmon:1996:DWD:240455.240470} and complements the data
management elements of existing digital curation models such as the Digital
Curation Centre (DCC) Lifecycle Model~\cite{higgins2008}.
Because we favor automatisation when possible, our approach is
founded on \emph{sheer curation}~\cite{macdonald}:  the curation 
activities are integrated  within
the normal workflow of those creating the data.

We review the ETL model in \S\ref{sec:etl} and show how it can be adapted to
the problem of extracting data in \S\ref{sec:e};
transforming it in \S\ref{sec:t} and archiving it in \S\ref{sec:a}.

\section{ETL}
\label{sec:etl}

ETL is a process model used in data warehousing  to integrate heterogeneous data sources and enable uniform data analytics. The ``Extract'' component of the
ETL process aims at harvesting data from disparate sources in a variety of formats. The ``Transform'' part of the ETL process performs cleaning operations
and applies encoding rules to convert the source data into a more coherent form. The ``Load'' phase takes the transformed data that conforms to a uniform data schema and makes it available to a database system on which, for example,
analysis tools can be executed.

In the Enterprise Database marketplace,  software tools, such as Oracle Warehouse Builder, DB2 Warehouse Edition and Microsoft SQL Server Integration Services implement and automate the ETL model. These tools pay particular attention to the enterprise needs of
performance and scalability as well as the requirements for data migration and auditing.

The parallels with the requirements for managing scientific research data on a large scale
are clear: research data, even within the same scientific discipline manifests in a variety
of heterogeneous formats and there is a present need for ``data harmonization''. The scale and distribution of research-data also means that the models and methods used in pure data warehousing should apply.

\section{Data Extraction}
\label{sec:e}

In their study of the life cycle of e-Science data, Wallis et
al.~\cite{1255228}  
identified the following
phases: (1)~Experimental Design, (2)~Calibration, (3)~Data Capture or
Generation, (4)~Data cleaning and Derivation, (5)~Data Integration, (6)~Data
Derivation, (7)~Data Analysis, (8)~Publication, Storage and Preservation. They
found that scientists need to access their data at each phase and must be able
to use and integrate data from multiple sources. As the authors point out

\begin{quote}
The lack of an integrated framework for managing these types of scientific data
presents significant barriers not only to those scientists conducting the
research, but also to those who would subsequently reuse the data.
\end{quote}

To enable reuse, we must publish the data.
Until the Web became ubiquitous, the data itself was rarely published
separately from the research articles, and for a good reason: data cannot be
understood without a context. Yet decoupling the data from its context 
is invaluable because it enables verification, reuse and re-purposing.

There are several strategies to decouple the data while retaining its link to its context.
The most prevalent approach is to require researchers to upload
their data to a curated repository after publication of the corresponding
research articles. For example, DiLauro et al.~\cite{1816200} describe a system
wherein the data is captured during the submission of the research article.
This ensures that the data is properly linked to the research article and that
data submission is part of the  researchers' workflow.

A preferable alternative is to systematically archive
the data as it is being collected and processed~\cite{cragin2010data}, or
even as it is being acquired by instruments such as some data repositories in
astronomy do when the data is collected by telescopes.
Finally, the data can be extracted from  
the research articles or reports themselves: from the tables of results, 
from the results section---commonly found in the abstract in medical
articles---or elsewhere in the document. 

Besides the data itself, we must also capture metadata
to help users retrieve, assess and reuse the data. 
Decoupled data may then be linked to a region of text---such as the text
which surrounds a table in a research article. Such text serves
as indexable metadata, as HTML text does for images on web indexes.

\subsection{Extraction from Legacy Sources}

Useful data may be inconveniently embedded in a variety of previously published
documents. For example, until recently, it was common to store data on paper as
plots. Thus, researchers are now forced to recover data by scanning plots
from research papers~\cite{mayersohn1999reclaiming}. 

Research data is also published as tables in PDF, HTML or XML documents. 
Thankfully, automated data extraction systems such
as Tableseer and the SciVerse Applications platform allow
researchers to search for and extract tables embedded within documents. 
Other publishers---such as the Public Library of
Science (PLoS)---make available the content of all their journals in XML, a
machine readable format that makes it convenient to extract data from published
articles using an XQuery engine.
Hence, massive numbers of documents can be processed automatically with little
effort.

\subsection{Decoupled and Linked Data}
Decoupled data  needs to meet only three criteria:
\begin{itemize}
\item It must be free from the confines of the publication format 
of a research article (PDF, DOC, HTML). That is, it should
be in a data-appropriate format that enables further machine processing (CSV, XML or SQL assertions). 
\item It must be reusable. Thus, it must be available, complete, licensed for reuse
and documented. If appropriate, it should follow known data models and schemas~\cite{1378912}.
\item It must be possible to refer to it to specifically and independently of any research article. For example,
it could have a unique identifier.
\end{itemize}
There is a distinction but also a relationship between the concepts of ``linked
data'' and ``decoupled data''. Linked data  exposes, shares, and connects
pieces of data using URIs and RDF~\cite{bizer2009linked}.
Hence decoupled data may become linked data. Indeed, decoupling the data from its textual source makes subsequent linking possible.

\section{Transformation}
\label{sec:t}

Scientists usually transform their data before archiving it:
\begin{description}
\item[Mergers and joins]
Researchers routinely integrate data sets from different sources to derive
indicators and measures: astronomers may combine the data from several
telescopes and geophysicists may combine satellite data with ground sensors.
A frequent, but mostly implicit, type of join
occurs when mostly static data is used as part of a derivation. For example, physical constants or geographical data is often used in conjunction with
recently collected data. 

\item[Data cleaning]
Almost all research data requires cleaning. The collected
data might be inconsistent or contradictory. Outliers indicating
faulty measures are common. 
A particular challenge in science is 
baseline correction. For example, climatologists need
to correct the temperature records for the effect
of growing cities. 

\item[Data filtering and aggregation]
It is common for scientists to record more data points
than needed: this extra data must be either filtered out or aggregated.
When medical researchers process electrocardiograms (ECGs), they 
routinely keep as little
as only the location of one data point per heart beat (e.g., the 
location the R wave).
Geophysicists may carry aeromagnetic survey using planes that
record several samples per second, whereas they are ultimately only
interested in a geological map having a relatively low resolution.

\item[Data mapping]
A common mapping in science is a change of units 
(e.g., from inches to cm). Numerical data can be rounded (e.g.,
to 3~significant digits). 
\end{description}

This list is by no means exhaustive.  Other transformations include
compression, deduplication and  validation~\cite{altman2008}. 
Moreover, scientists increasingly work with data sets so large that
they cannot manually inspect them. We must rely on algorithms.
Thankfully, there are user-friendly tools to help users transform
their data more reliably~\cite{cyganiak2010}. In this respect, we find
Google Refine~\cite{cyganiak2010} particularly interesting.

\subsection{Formatting and Standardization}

An important type of transformation is the one that maps the data between different formats.
For example, long term archival may require a machine-independent format such
as netCDF~\cite{rew2002netcdf} whereas, for  on-line access, it might be
preferable to have the data in an SQL format.

Beyond the data tself, the metadata must also be properly formatted for
interoperability and long-term storage. For example, the Core Scientific
Metadata Model  (CSMD)~\cite{matthews2010using} is generic enough to apply to a
variety of disciplines but also detailed enough to enable the reuse and
repurposing of data within and across scientific disciplines.

\section{Archiving}
\label{sec:a}

We distinguish three types of data which may require archival: 
\squishlist
\item Raw data, which might result directly from an experiment or a simulation, or it might have been extracted from legacy sources.
\item Derived data, which is the result of any processing on the raw data,
including cleaning (correction for errors). It includes data integration
wherein various data sets are used to create a new data set.
\item Resultant data, which is the final product, typically what might be
published by the authors along with their research article.
\squishend

After several decades of manual curation, scientific data repositories such as
GenBank~\cite{benson1998genbank} offer a wealth of raw data  and associated
metadata, including references to the published and gray literature. There are
even journals such as Earth System Science Data  dedicated to publishing raw
data. 

We know from this experience that a proper data archive must support data
embargoes~\cite{cragin2010data} and must provide access control. This is
especially of concern if researchers upload their data prior to the final
publication of their research articles. Sometimes the data needs to be archived
and accessible while remaining  partially confidential.

Moreover, a data archive should support versioning: even within a single team,
there might be several versions of the same data set~\cite{wallis2008}. For
long-term storage, data must be protected against loss and corruption as
well as malicious attacks~\cite{1555457}. Data sets should be properly indexed
and documented and should have unique identifiers.

While it might be tempting to only store the resultant data, 
there are at least two problems with this approach: 
\squishlist
\item other researchers may mistrust transformations that they cannot verify;
\item  it is difficult to predict how and in what format the data might be most
useful to others, even with the best intentions.
\squishend
Thus, publishing only the resultant data may limit its
reusability~\cite{Malik2010}. Moreover, as Yan et al.~\cite{Yang2010} report,
``scientists are highly motivated to publish the entire data trail along the
analysis pipeline.''

Data archiving with an ETA framework affords an opportunity to do more than mere curation. The
association of research data with other artefacts such as research articles
makes it possible to automate the analysis of metadata to discover trends in
the published literature. Thus it should be possible to measure whether
progress toward knowledge objectives have been achieved. Similarly, such
metadata analysis could detect anomalies and inconsistencies among research
results. Citation data analysis could be the basis for recommending research
data sets and research papers~\cite{Sugiyama:2010:SPR:1816123.1816129} to
researchers. Last but not least it is also possible to mine the data itself to
discover novel results~\cite{macmillan2009nextbio}.

\subsection{Specialization or Integration?}

In conventional data warehousing, experts distinguish between data marts, which
are specialized domain-specific data repositories (e.g., for accounting) and
integrated data warehouses, which provide a uniform layer of abstraction from
the data-domain. In the context of scientific data repositories, the Australian
National Data Service (ANDS) is an example of the later whereas GenBank is an
example of the former.

While domain-specific repositories such as GenBank are simpler to setup and
maintain, they are not as conducive to interdisciplinary research. Even though
they are more difficult to implement, integrated data warehouses afford a greater
likelihood for the data to be repurposed across different disciplines. They are more
likely to persist over time and are easier for machines to resolve outside of
the subject domain context, such as the database system that generated them.
An example of this difference is found in the data-identification mechanism.
GenBank assigns its own unique dataset identifier (accession numbers) based on
subject-domain conventions for referencing genes and proteins. However, for
interdisciplinary research, it might be preferable to refer to those same
datasets with a domain-independent persistent identifier system such as DOIs
for data sets~\cite{treloar2008access}  
granted by institutions
affiliated with DataCITE~\cite{brase2009}. One advantage of adopting a 
domain-independent identifier is that it eliminates the need for multiple methods for
name-resolution.

The trade-off between the two methodologies (centralized versus specialized)
is well documented in the data warehousing
literature~\cite{Jukic:2006:MSA:1121949.1121952}. Experience suggests that
the integration of data marts may lead to metadata inconsistencies while the
integrated data warehouse approach is costly and difficult to initiate.
Specialized data marts tend to be more concise as only the information
deemed relevant by the community is included. Integrated data warehouses are
more likely to rely on homogeneous technology: they use fewer software
vendors. Data marts are often more dynamic: it is comparatively easier to
add new feature or new metadata when you must only address the needs of a
specific community.

\subsection{A Diverse Software Architecture}

Many bibliographic repositories---of either text or data---suffer from a common ailment that could be addressed  within our proposed ETA process.  
Indeed, data archiving is often performed by relational databases
whose core concepts were invented in the 1960s, 
and whose technology is insufficient
with respect to modern information retrieval needs: semantic search, question answering, content-clustering and dynamic schemas,  to name only a few.
For these purposes, full-text indexing tools such as Apache Lucene that can do full text analysis e.g., stemming, part-of-speech tagging, term-frequency analysis are starting to replace databases.
Similarly, document-oriented databases such
as CouchDB and MongoDB might offer the necessary flexibility to dynamically support many different
database schemas corresponding to different domains, with relatively 
little maintenance.

\section{Conclusion and Future Research}

The management of scientific data repositories can benefit from lessons learned in data 
warehousing~\cite{Inmon:1996:DWD:240455.240470}.
Both specialized data marts and integrated data warehouses have a role
to play in the data archiving ecosystem. When integrating heterogeneous data
is too difficult, the data mart approach is preferable. Otherwise, An
integrated data warehouse approach favours interdisciplinary collaboration: it
offers uniform metadata conventions, persistent identification nomenclature
and better automatisation for ingest. However, if the sources of heterogeneous
data are too diverse, the domain-specific data mart approach may be preferable.
Moreover, the choice might be guided by the available funding: in most instances,
the integrated approach will prove more expensive and require more time.

In either case, data cannot be routinely processed in an ad hoc fashion.
The ETA process must be automated as much as possible. While some data
repositories can develop their own automation architecture, there is an
opportunity to develop more generic ETA tools. For example, we could
extend existing open source ETL tools such as Talend\footnote{\url{http://www.talend.com/}} 
or Pentaho Data Integration (PDI)\footnote{\url{http://www.pentaho.com/products/data_integration/}}.

Donoho et al.~\cite{donoho2009reproducible} recommend making available both the data and 
the instructions necessary to reproduce any published figure or other published objects. 
While journals and funding agencies may require that the data is available,  as yet we lack
conventions on how to document and archive the transformation from raw data to, for example, a figure. 
Ideally, the data used to generate a figure or a table should always be
available through a permanent identifier such as a Data DOI.

Extracting, transforming and archiving heterogeneous data can
accommodate a diversity of software architectures. Expertise in data
warehousing require the ability to integrate a wide variety of technologies,
data formats and data models. We cannot expect to index all research data
using only a few simple models: data archives must embrace diversity. To cope
with this diversity, we need extensible data-management tools.

An essential insight is that managing the flow of data, from its extraction to
its storage and retrieval, is often more important than merely curating the
provided data~\cite{Yang2010}: the life-span of raw data may also include
derived and resultant data. Furthermore, it is necessary to consider that data
may originate from extraction processes and appropriately transformed and
identified.

%
%
\bibliographystyle{abbrv}
\bibliography{ETA}

\end{document}